\documentclass[preprint,aps]{revtex4}

\usepackage{amsmath}
\usepackage{amssymb}

\usepackage{epsfig}
\newcommand{\lbfig}[1]{\refstepcounter{fig} \label{#1} }
\newcounter{fig}

\begin{document}

\preprint{ITP-UU-06/53, SPIN-06/43}

\centerline{\Large \bf  Baryogenesis in Brans-Dicke theory}

\author{Sietse van der Post}
\email[]{post@phys.uu.nl}
\author{Tomislav Prokopec}
\email[]{T.Prokopec@phys.uu.nl}
\affiliation{Institute for Theoretical Physics (ITP) \& Spinoza Institute,
             Utrecht University, Leuvenlaan 4, Postbus 80.195, 
              3508 TD Utrecht, The Netherlands}

\begin{abstract}

A new mechanism for baryogenesis is proposed in the
context of an extended Brans-Dicke (BD) theory. We generalize the BD scalar
to complex field with CP violating coupling to curvature and 
show that the charged BD current can be enhanced during inflationary epoch. 
After inflation the current decays (via tree level
interactions) into the standard model particles. When the BD scalar  
is charged under baryon and/or lepton number, the decay 
produces a net baryon number. Rather generically a sufficiently large
scalar current can be produced during inflation to account
for the observed baryon asymmetry of the Universe. 
Our baryogensis scenario can in an elegant
way be incorporated into a model of extended inflation.

\end{abstract}

%\keywords{}

%98.80.-k Cosmology 
% (see also section 04 General relativity and gravitation; 
% for elementary particle and nuclear processes, see 95.30.Cq; 
% for dark matter, see 95.35.+d; 
% 95.36.+x Dark energy
% 98.80.Cq Particle-theory and field-theory models of the early Universe
% (including cosmic pancakes, cosmic strings, chaotic phenomena,
%  inflationary universe, etc.)  

\maketitle

\section{Introduction}

In an expanding universe particle densities scale rather generically
as $n\propto 1/a^3$, where $a$ denotes the scale factor of the Universe.
This observation immediately implies that any charge density
preceding the inflationary epoch is diluted during inflation into oblivion.
Namely the volume during inflation expands so tremendously
that information about any initial charge is forgotten
during inflation. This is true of course only if the dilution is not 
compensated by the production. To our knowledge up to now 
no mechanism has been proposed by which a charge density in some particles
would grow during inflation. Here we propose such a model within 
the context of an extended Jordan-Fierz-Brans-Dicke (BD) 
theory~\cite{Brans:1961sx},
where the BD scalar is generalised to a complex scalar field. 
We shall argue that under rather generic conditions, 
an initial small scalar charge density extending
over several Hubble volumes can be efficiently amplified during inflation.
 At the end of inflation the charge decays into ordinary matter
fields, converting thus the scalar charge into baryonic or/and leptonic 
charges, which is in turn reprocessed by sphalerons into a baryon asymmetry,
representing thus a mechanism for creation of the
observed baryon asymmetry of the Universe.

 The use of scalar fields has been ubiquitous in baryogenesis models, 
ranging from Affleck-Dine baryogenesis~\cite{Dine:2003ax,Affleck:1984fy}, 
spontaneous baryogenesis~\cite{Cohen:1988kt}, 
to electroweak scale baryogenesis~\cite{Cohen:1991iu,Huber:2006wf}. 
Yet with a few exceptions~\cite{Dolgov:1991fr,Balaji:2004xy}, 
inflation has not been used in connection with baryogenesis.

The attractive feature of our baryogenesis model
is that the field responsible for the strength 
of the gravitational coupling is also responsible for the generation
of matter. In General Relativity {\it no} such
mechanism is possible since the metric tensor cannot violate CP symmetry.

The plan of the Letter is simple. 
In section~\ref{Brans-Dicke theory with a complex scalar}
we define our generalized Brans-Dicke theory. 
The resulting equations are solved and analysed in 
section~\ref{Analysis of the model} and 
in section~\ref{Conclusion} we conclude.

\section{Brans-Dicke theory with a complex scalar}
\label{Brans-Dicke theory with a complex scalar}

 In Brans-Dicke theory the gravitational constant $G$ is promoted to 
a dynamical parameter changing in space and time \cite{Brans:1961sx}. 
This variation is conventionally expressed through a scalar field $\Phi$ 
such that $G=G_0 \Phi^{-1}$ with $G_0$
the bare, true constant of nature. The Brans-Dicke action is in
natural units ($c=G_0=\hbar=1$) given by,
\begin{equation}
  S=\frac{1}{16\pi}
      \int d^4 x (-g)^{1/2} \left[ \Phi R - \frac{\omega}{\Phi}(\partial_{\mu} \Phi \partial_{\nu} \Phi)g^{\mu \nu}
      \right]+\int d^4 x {\cal L}_{\rm mat}
\,.
\label{standbd}
\end{equation}
Here $R$ denote the Ricci scalar curvature, ${\cal L}_{\rm mat}$ is the
ordinary Lagrangian of matter and non-gravitational fields,
$\omega$ is the BD coupling constant and $g={\rm det}(g_{\mu\nu})$.
For dimensional reasons as
well as for convenience the substitution $\Phi \rightarrow \phi
\phi$ is sometimes made in literature. We will make an analogous
substitution, since it converts the kinetic term to the canonical
form. 

 Here we propose the following generalisation of the Brans-Dicke theory.
We promote $\phi$ to complex scalar
and write the scalar sector of the theory~(\ref{standbd}) as follows, 
\begin{eqnarray}
(-g)^{-1/2}{\cal L}_\phi = -\rho(\partial_{\mu}\phi
      \partial_{\nu}\phi^*)g^{\mu\nu}
    + \Big(
       -\frac12\bar\omega(\partial_{\mu} \phi \partial_{\nu} \phi)g^{\mu\nu}
       + \bar\mu\phi^2 R
    + {\rm h.c.}\Big)
\,.
\label{our lagrangian}
\end{eqnarray}
Note that this lagrangian contains operators up to dimension four only,
and in this sense the renormalisability properties of gravity are 
not made worse by the theory~(\ref{our lagrangian}).
Furthermore, it can accomodate CP violation when 
the (dimensionless) parameters $\bar\omega$ and $\bar\mu$ are complex,
\begin{equation}
  \bar\omega = \omega {\rm e}^{i\theta_\omega}
,\quad (\omega \equiv |\bar\omega|)
\,,\qquad
  \bar\mu = |\bar\mu| {\rm e}^{i\theta_\mu}
\,.
\label{complex couplings}
\end{equation}
 There is only one physical CP-violating phase in Eq.~(\ref{our lagrangian}),
however. Indeed, the phase of $\bar\omega$ can be removed by a field rotation, 
$\phi \rightarrow \phi\, {\rm exp}(-i\theta_\omega/2)$, upon which 
the kinetic terms in Eq.~(\ref{our lagrangian}) become real. 
The remaining complex parameter,
\begin{equation}
 \mu \equiv \bar\mu {\rm e}^{-i\theta_\omega} 
        = \mu_r + i \mu_i
\,,
\label{bar_mu}
\end{equation}
whose components are of the form, 
\begin{equation}
   \mu_r  = |\mu| \cos(\theta_\mu-\theta_\omega)
\,,\qquad
   \mu_i = |\mu| \sin(\theta_\mu-\theta_\omega)
\,,
\label{bar_mu:r+i}
\end{equation}
mediates CP violation of the theory~(\ref{our lagrangian}). 
Note in particular that, when 
$\theta_\mu-\theta_\omega \neq n\pi$ ($n=$integer), then $\mu_i \neq 0$,
and there is a nonvanishing CP violation in the theory. 
Moreover, when $\pi/2 < \theta_\mu-\theta_\omega < 3\pi/2$, then $\mu_r<0$.
As we shall see, the case of negative $\mu_r$'s 
plays an important role in the analysis that follows.

 Dubbing the real and imaginary component of the rotated field
as $\phi_{\pm}$ such that,
\begin{equation}
 \phi {\rm e}^{-i\theta_\omega/2} = \phi_+ + i\phi_-
\,,
\label{phi:+-}
\end{equation}
we can rewrite lagrangian~(\ref{our lagrangian}) as,
\begin{eqnarray}
(-g)^{-1/2}{\cal L}_\phi 
  = -(\rho+\omega)(\partial_{\mu} \phi_+ \partial_{\nu}\phi_+)g^{\mu \nu} 
 - (\rho-\omega)(\partial_{\mu} \phi_- \partial_{\nu} \phi_-)g^{\mu\nu}
 + 2\mu_r(\phi_+^2-\phi_-^2)R - 4\mu_i\phi_+\phi_- R
\,.
\label{kinlag}
\end{eqnarray}
 The coefficients of the kinetic terms have to be such that
$\rho-\omega>0$ in order to have positive energy states. As can be
seen the original lagrangian corresponds to two fields with
different dynamics. The presence of charge violation should be
clear by the mixing between the two fields in the last term and
the sign difference in the quadratic terms. We emphasize that, 
although this model can be written in terms of two coupled
real scalars, the physics is different. The two fields
are connected through the relation 
$\phi {\rm e}^{-i\theta_\omega/2}=\phi_+ +i\phi_-$,
such that a non-vanishing scalar current can arise.

 Since in our model baryon number is created via a postinflationary decay of 
a scalar current $j_\phi^0 \equiv Q$, 
we shall mainly be interested in the production of a scalar charge density 
$q_\phi$ during inflation, 
\begin{eqnarray}
q_\phi \equiv J_{\phi}= \frac{i}{2}(\phi\dot{\phi}^* - \phi^*\dot{\phi})
% = \phi_+\dot{\phi}_- - \phi_-\dot{\phi}_+
% = \frac{1}{2\sqrt{\rho^2-\omega^2}}
%     \left(\psi_+\dot{\psi}_- - \psi_-\dot{\psi}_+\right)
\,.
\label{cur}
\end{eqnarray}
A scalar field is said to be charged under the baryon number B 
(or any other charge not proportional to B+L)
if -- when it decays via tree level processes into the standard model 
matter fields -- it generates a net lepton an baryon number
according to the relation, 
\begin{equation}
   Q = q_\phi V = q_B B + q_L L 
\,,
\label{Q}
\end{equation}
where $Q$ denotes the total scalar charge, $V$ is the spatial volume,
$B$ and $L$ denote the baryon and lepton number densities, respectively, 
and $q_B$ and $q_L$ are the corresponding charges. 
Sphalerons~\cite{Manton:1983nd,Klinkhamer:1984di} 
then process any $(B-L)_0$ charge produced during 
inflation into a net $B$ and $L$ according to 
\cite{Harvey:1990qw},
\begin{eqnarray}
  B &=& \frac{8n_f + 4(n_H+2)}{24n_f + 13(n_H+2)}(B-L)_0
\nonumber\\
  L &=& - \frac{16n_f + 9(n_H+2)}{24n_f + 13(n_H+2)}(B-L)_0
\,,
\label{B final}
\end{eqnarray}
where $n_F$ and $n_H$ denote the number of quark and lepton families and 
Higgs doublets, respectively. 
This can be then observed as the baryon and lepton asymmetry of 
the Universe today.

\section{Analysis of the model}
\label{Analysis of the model}

We shall now analyse the dynamics of the scalar field
governed by the lagrangian~(\ref{our lagrangian}) (and~(\ref{kinlag}))
during inflation. 
 Inflation in a Brans-Dicke theory~\cite{La:1989za} is of a powerlaw type
with the Ricci curvature scalar decaying as $R\propto t^{-2}$. 
The solutions for the scalar $\Phi$ and corresponding scale factor in the
original BD theory of Eq.~\eqref{standbd} are,
\begin{align}
\Phi & = M_p^2(1+\chi t/\alpha)^2 
\nonumber\\
  a(t)& =a_0(1+\chi
t/\alpha)^{\omega+1/2}
\,,
\label{BD:phi-a}
\end{align}
where $M_p = (8\pi G)^{-1/2}$
denotes the Planck mass, $\chi^2 = 8\pi \rho_f/3M_p^2$
the Hubble constant squared in the Einstein frame,
$\alpha^2 = (3+2\omega)(5+6\omega)$, and 
$\rho_f$ is the energy density driving inflation.
As can be seen, for early times
and large $\omega$ the expansion is nearly exponential
$a(t)\approx e^{\chi t}$ and $R$ is decays slowly. 
For simplicity here we assume that
$\omega$ in Eq.~(\ref{BD:phi-a}) is large, such 
that we can approximate the space-time by de Sitter space 
with the metric tensor given by
\begin{equation}
 g_{\mu\nu} = {\rm diag}(-1,a^2,a^2,a^2)
\,,\qquad a = a_0 \exp(Ht)
\,,
\label{metric tensor:dS}
\end{equation}
where $a$ denotes the scale factor
($a_0$ is some initial scale factor of no physical relevance)
and $H$ is the Hubble parameter.

 In de Sitter space~(\ref{metric tensor:dS}) 
the lagrangian~\eqref{kinlag} implies the following equation of motion
for the homogeneous modes $\phi_\pm=\phi_\pm(t)$,
\begin{eqnarray}
\ddot{\phi}_{\pm} + 3H\dot{\phi}_{\pm}
                 \mp 2\xi_{\pm}R\phi_{\pm}
                  + \zeta  R\phi_{\mp}
                  = 0
\,,
\label{eomkin}
\end{eqnarray}
where 
\begin{equation}
 \xi_{\pm} = \frac{\mu_r}{\rho \pm \omega}
\,,\qquad
 \zeta = \frac{2\mu_i}{\sqrt{\rho^2 - \omega^2}}
\,.
\label{xi+-:zeta}
\end{equation}
Note that we are considering the evolution of charged  
configurations which are to a good approximation homogeneous 
at the beginning of inflation. If the Universe was highly 
inhomogeneous at the beginning of inflation, only a small fraction 
of field configurations qualify, such that the corresponding 
initial amplitude may be small~\cite{Balaji:2004xy,Dolgov:1991fr}.
The amplitude of these large scale charge fluctuations, which we 
assume to be of statistical origin, can be estimated as follows. 
The number of particles $N$ and antiparticles 
$\bar N$ in a volume $V\sim H^{-3}$ are given by $N \sim T^3 V$,
$\bar N \sim T^3 V$, and the corresponding charge is $Q\sim N-\bar N$, where 
$T$ denotes the effective temperature (the system is not assumed to be
in thermal equilibrium). To estimate the amplitude of charge fluctuations
$\delta Q$, note that because of total charge neutrality of the system,
each charge is paired with an opposite charge, such that 
the boundary surface $A = \partial V$ contains
$N_A \sim  A/\ell^2 \sim T^{2} V^{2/3}$ charge pairs,
where $\ell \sim T^{-1}$ is the average charge separation.
 The statistical charge fluctuation
$Q_A = \sqrt{N_A}$ then corresponds to the expected net charge 
in the volume $V$, $\delta Q \sim Q_A \sim T V^{1/3}$. Given that 
$V^{1/3}\sim 1/H$ and $T\sim \sqrt{HM_P}$, we get for the charge fluctuation 
in a Hubble volume $\delta Q \sim (M_P/H)^{1/2}$, which gives for the initial 
current, $j_\phi \sim \delta Q/V\sim H^3 (M_P/H)^{1/2}$. 
The charge fluctuation $\delta Q$ is of course suppressed
when compared with the total particle number 
$N\sim (M_P/H)^{3/2}$ as $\delta Q/N = j_\phi/(N/V) \sim H/M_P$. 
For example, when this estimate is applied to the electroweak scale, one gets 
for the baryon to entropy ratio, 
  $n_B/s \sim (T/M_P)^2 \sim H/M_P \sim 10^{-34}$.
 From this estimate and the analysis presented below it then follows that
-- as long as the charge does not decay during inflation --
the preinflationary charge density fluctuations should  
suffice to make our baryogenesis mechanism effective.  

 In de Sitter space~(\ref{metric tensor:dS}) 
Eq.~(\ref{eomkin}) can be solved by an exponential {\it Ansatz}. 
Since the equations are coupled, there are four independent solutions
with corresponding real constants $\phi_{\pm j}^{(0)}$.
The solutions read,
\begin{eqnarray}
 \phi_{\pm} &=& \Re\sum_{j=1}^4 \phi_{\pm j}^{(0)} e^{\kappa_j t}
,\quad\quad  
\kappa_{\pm \pm}=-\frac32 H \left(1\pm\sqrt{1-\beta \pm \gamma}\right)
\\ \nonumber 
 \beta &=& \frac{8R\omega\mu_r}{9H^2(\rho^2-\omega^2)},
\quad\quad
 \gamma = \frac{8R}{9H^2(\rho^2-\omega^2)}
          \sqrt{\rho^2\mu_r^2 + (\rho^2-\omega^2)\mu_i^2}
\,.
\label{soln}
\end{eqnarray}
Note first that only four out of the eight constants
$\phi_{\pm j}^{(0)}$ are independent. This is so because 
$\phi_{+j}^{(0)}$ and $\phi_{-j}^{(0)}$ are related by Eq.~(\ref{eomkin})
as follows, 
\begin{equation}
 [\kappa_j(\kappa_j+3H) - 2\xi_+R]\phi_{+j}^{(0)} + \zeta R\phi_{-j}^{(0)} = 0
\,.
\label{fields:condition}
\end{equation}
Secondly, as can easily be verified, only $\kappa_{--}$ and $\kappa_{-+}$
can be positive and hence can give growing
solutions. For certain choice of parameters $\kappa_{--}$ can
become complex, though. In order to have a growing current, we
need two independent solutions, since the contribution of
solutions of $\phi_{\pm}$ with equal time behavior is zero as can
be seen from Eq.~\eqref{cur}. The leading contribution
therefore comes from the interference between $\kappa_{--}$ and
$\kappa_{-+}$. Neglecting the other two decaying terms, we adopt
the solution,
\begin{equation}
\phi_{\pm}=\Re\left[ \phi_{\pm 1}^{(0)} e^{\kappa_{--}t} + \phi_{\pm 2}^{(0)}
             {\rm e}^{\kappa_{-+}t} \right],
\end{equation}
where we took the real part since $\phi_{\pm}$ are by definition
real. Substituting this in definition~\eqref{cur} of the current
we arrive at,
\begin{align}
J_{\phi}=(\phi_{+1}^{(0)} \phi_{-2}^{(0)}-\phi_{+2}^{(0)} \phi_{-1}^{(0)})
\Re\left\{\left(\kappa_{-+}-\kappa_{--}\right)
          {\rm e}^{(\kappa_{--}+\kappa_{-+})t}
   \right\}
%\left\{ \Re\left[{\rm e}^{\kappa_{--}t}\right]
%        \Re\left[\kappa_{-+}{\rm e}^{\kappa_{-+}t}\right]
%      - \Re\left[\kappa_{--}{\rm e}^{\kappa_{--}t}\right]
%        \Re\left[{\rm e}^{\kappa_{-+}t}\right]
%\right\}
\,,
\label{curkin}
\end{align}
where we took account of the fact that in order to get a growing 
solution, $\Re[\kappa_{--}+\kappa_{-+}]>0$, implying that 
$\kappa_{-+}$ must be real.

\bigskip

 There are two cases of interest:

\begin{itemize}

\item[] {\tt Case A.} When $1-\beta-\gamma <0$, then 
$\kappa_{--}$ is complex and $\kappa_{-+}$ is real 
with  $1-\beta+\gamma >4$, which implies:
\begin{equation}
  1-\gamma < \beta< \gamma-3
\,,
\label{condition A}
\end{equation}
from which we also infer that $\gamma>2$.
 In this case the current oscillates with 
a growing envelope amplitude. 
In order to prevent fast oscillations, some tuning 
is required. In particular, the oscillations are slow if 
$\beta \approx 1-\gamma$. In this case $\beta<0$ and 
$\mu_r<0$, since $\gamma>2$. The corresponding current is
\begin{equation}
 J_{\phi}=J_0^{(1)}\left[\cos(i\alpha_-Ht) +
         \frac{i\alpha_-}{\alpha_+} \sin(i\alpha_- Ht)\right]
  {\rm e}^{(\alpha_+-3)Ht}
\,,
%, \quad\quad
%    (1-\gamma < \beta< \gamma-3)
\label{current A}
\end{equation}
where 
\begin{equation}
\alpha_+ =\frac32\sqrt{1-\beta + \gamma}
\,,
\quad\quad
\alpha_- = \frac32\sqrt{1-\beta-\gamma}
\,,
\quad\quad
J_0^{(1)} = H\alpha_+(\phi_{+1}^{(0)} \phi_{-2}^{(0)}
           - \phi_{+2}^{(0)}\phi_{-1}^{(0)})
\,.
\label{alpha+-J2}
\end{equation}
A typical evolution of the
current~(\ref{current A}--\ref{alpha+-J2}) is shown in figure~\ref{figure 1}.
\begin{figure}[htbp]
\begin{center}
\epsfig{file=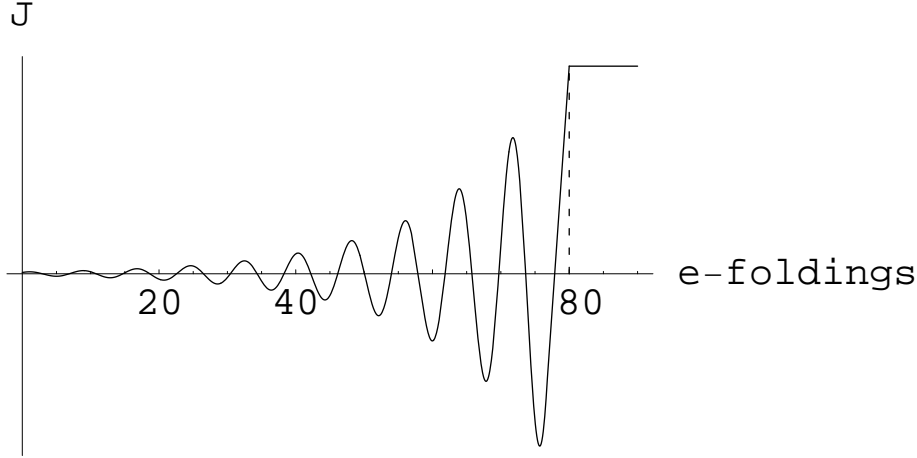, height=3.in}
\end{center}
\vskip -0.35in
\lbfig{figure 1}
\caption[fig1]{%
\small
 The scalar charge density (current) during inflation as a function of 
the cosmic time (the number of e-foldings). Note oscillations 
with a growing amplitude. After inflation, the charge density freezes out. 
}
\end{figure}

\item[] {\tt Case B.} When $1-\beta-\gamma >0$, both 
$\kappa_{--}$ and $\kappa_{-+}$ are real. In this case the current grows 
provided the condition
\begin{equation}
  \beta< - \frac{\gamma^2}{4}
\label{condition B}
\end{equation}
is fulfilled.
In this case no oscillations are present. From Eq.~(\ref{condition B})
we infer that $\mu_r<0$, and that there is a solution provided 
$|\mu_i| < 9H^2\omega\sqrt{\rho^2-\omega^2}/(4R\rho)$.
In this case the current grows exponentially,
\begin{equation}
J_{\phi} = J_0^{(2)} {\rm e}^{(-3+\alpha_++\alpha_-)Ht}
%(\beta < -\gamma^2/4) 
\,,
\label{current B}
\end{equation}
with
\begin{equation}
 J_0^{(2)} = (\phi_{+1}^{(0)}\phi_{-2}^{(0)}-\phi_{+2}^{(0)}\phi_{-1}^{(0)})
             H(\alpha_+-\alpha_-)
\,.
\label{J01}
\end{equation}
A typical evolution of the current~(\ref{current B}--\ref{J01}) is shown in 
figure~\ref{figure 2}.
\begin{figure}[htbp]
\begin{center}
\epsfig{file=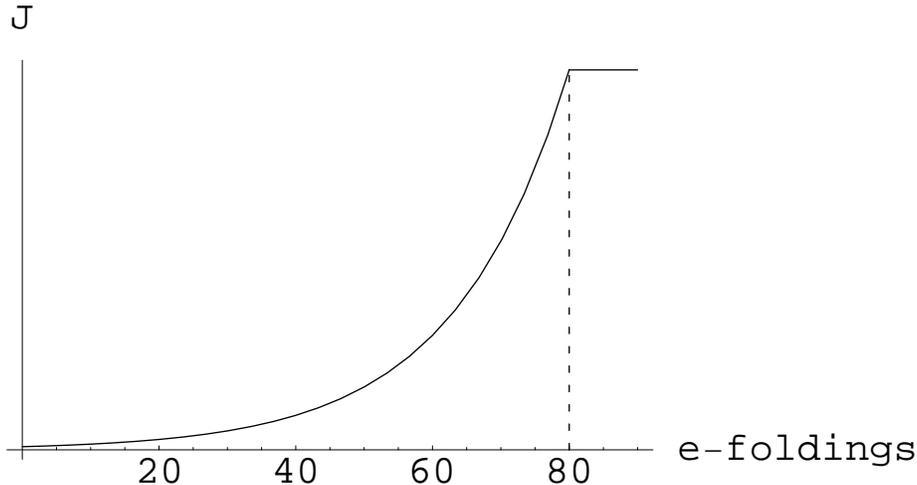, height=3.in}
\end{center}
\vskip -0.35in
\lbfig{figure 2}
\caption[fig1]{%
\small
 The scalar charge density during inflation. Unlike 
in figure~\ref{figure 1} no oscillations are present and the charge density
grows exponentially. 
}
\end{figure}

\end{itemize}

 By making use of Eq.~(\ref{fields:condition}) we can relate
the $\phi_{+1}^{(0)}$ and $\phi_{+2}^{(0)}$ such that  
the prefactor in Eqs.~(\ref{alpha+-J2}) and~(\ref{J01}) becomes,
\begin{align}
\phi_{+1}^{(0)} \phi_{-2}^{(0)}-\phi_{+2}^{(0)} \phi_{-1}^{(0)} 
 = -  \frac{4\phi_{+1}^{(0)}\phi_{+2}^{(0)}}{\mu_i}
     \sqrt{\frac{\rho^2}{\rho^2-\omega^2} \mu_r^2 + \mu_i^2}
\,.
\end{align}
Note that this relation and the above analysis are valid only in the presence 
of CP violation when $\mu_i \neq 0$. Indeed, Eq.~(\ref{eomkin}) 
implies that for $\mu_i=0$ the evolution of 
$\phi_+$ and $\phi_-$ decouple and $\phi_+^{(0)}$ and $\phi_-^{(0)}$ 
can be specified independently. In this case there is no CP violation in
the lagrangian. Nevertheless, even in this case an initial charge 
can be amplified provided there is a growing solution for the current.
The solution is however always oscillatory and oscillations are rapid.
On the other hand, amplification within a CP violating theory
is more efficient and -- as we have seen above --
for a certain choice of parameters amplification is purely exponential.  

  In the above analysis we have assumed that
$R$ is varying very slowly and expansion is close to exponential.
 Most inflationary models, including inflation driven 
by a Brans-Dicke scalar field, are nearly exponential (powerlaw). 
Our preliminary investigation of powerlaw inflation indications that
the current enhancement persists, albeit in that case it is weaker.

 An equivalent charge amplification can be obtained during inflation
with the replacement of $\mu^2R\rightarrow m^2$, with $m$ being 
the scalar mass. In that case CP violation remains operative after inflation,
resulting in nontrivial effects in radiation and matter epochs.
On the other hand, the curvature coupling is turned off 
in postinflationary radiation era, 
which makes the mechanism presented here in some sense simpler. 
Moreover, in the massive scalar case, no clear connection to the
gravitational sector exists. Thus unless one finds another motivation 
for the scalar field (like e.g. flat directions in supersymmetric affleck-Dine
baryogenesis models~\cite{Dine:2003ax,Affleck:1984fy}), 
the main impetus for this scalar-field baryogenesis mechanism would be lost. 

 After inflation the Universe enters a preheating epoch. 
The charged current can decay through a tree level coupling to ordinary matter
in a charge conserving manner. 
Since the Brans-Dicke scalar is not charged under the gauge group 
of the Standard Model, it cannot 
couple directly to charged fermions and leptons of the Standard Model.
It can however couple to neutrinos~\cite{Balaji:2004xy} {\it via}
the Yukawa-type lagrangian,
\begin{equation}
 {\cal L}_{\rm int} \supset - \sum_{i,j=1}^3 ( \bar\nu_R^i y_{ij}\phi \nu_L^j
                       + {\rm h.c.} )
\,.
\label{interaction lagr}
\end{equation}
Note that when $\phi$ acquires an expectation value,
this lagrangian generates Dirac neutrino masses.
The interaction~(\ref{interaction lagr})
 induces the following tree level decay processes
\begin{equation}
  \phi \rightarrow \bar\nu_L^i+\nu_R^j
\,,\qquad 
  \phi^* \rightarrow \bar\nu_R^i+\nu_L^j
\,,
\label{tree decays}
\end{equation}
with the coupling strengths $y_{ij}$ and $y_{ij}^*$, respectively, such 
that upon decay the scalar charge is converted into a lepton charge
of a similar amplitude
\begin{equation}
J_{\phi} = Q \simeq L_0
\,.
\label{lepton charge}
\end{equation}
When sphalerons become active (below about $T\sim 10^{12}~{\rm GeV}$, which
corresponds to the sphaleron equilibration temperature), 
the lepton charge~(\ref{lepton charge}) gets converted
into a net baryon (and lepton) number according to Eq.~(\ref{B final}),
with $(B-L)_0 \equiv -L_0$. For example, for $n_f = 3$ and 
$n_H = 1$, Eq.~(\ref{B final}) gives $B = - (8/37)L_0 \simeq 0.2 Q$. 

 During preheating a lot of entropy is produced, which results in 
an entropy density of the order $s\sim T^3$.
Assuming instant reheating, the reheat temperature is
$T\sim \sqrt{HM_P}\sim V^{1/4}$, such that the entropy 
density is at most, $s\sim V^{3/4}$.
From this and  the observed baryon-to-entropy 
ratio \cite{Bennett:2003bz,Spergel:2006hy},
\begin{equation}
\frac {n_B}{s}
 %\equiv \frac{n_b - n_{\bar{b}}}{s} 
 = (8.7\pm 0.3) \times 10^{-11}
\,,
\end{equation}
we conclude that the required scalar current at the end of inflation is 
of the order 
\begin{equation}
 J_\phi \sim 10^{-9} V^{3/4}
\,,
\end{equation}
where 
$V \sim (10^{16}~{\rm GeV})^4$ is the potential energy driving inflation.
A comparison with the entropy density at the end of inflation,
which is of the order $s_H \sim H^3$, is also instructive. 
The entropy production during preheating is of the order, 
$s_T \sim V^{3/4} \sim (V^{1/4}/H)^3 s_H \sim (M_P/H)^{3/2} s_H$,
which is in typical one field scalar inflationary models of the order
$10^8 s_H$. In these models we therefore need
to produce a scalar charge density at the end of inflation of the order, 
\begin{align}
\left. J_{\phi}\right|_{\text{end infl}} \sim 10^{-1} H^3
\,.
\label{postinflationary amplitude}
\end{align}
When compared with the above estimate of the expected magnitude of scalar 
charge fluctuations in the primordial preinflationary chaos,
the required postinflationary current~(\ref{postinflationary amplitude}) 
seems easily feasible within the mechanism presented here.

One of the merits of the model is that it fits well into a model
of extended inflation. Equation~\eqref{kinlag} resembles the
lagrangian of curvature coupled extended inflation as proposed by Laycock and
Liddle \cite{Laycock:1993bc}, except for the addition of the mixing
term due to CP violation and the absence of an inflaton potential.
Depending on the sign of $\mu_r$, 
the fields $\phi_\pm$ assume the role of the inflaton or of the BD scalar. 
An example of the inflationary potential in which the inflaton couples  
to the Ricci scalar is,
\begin{align}
V(\phi)=&\lambda(|\phi|^2-|\phi_0|)^2 =
\sum_{\pm}\lambda(\phi_{\pm}^2-|\phi_0|^2)^2 + 2\lambda \phi_+^2
\phi_-^2 
\label{powerlaw inflation:V}
%\\ =&
%\frac{\lambda}{(\rho+\omega)}(\phi_+^2-|\phi_0|^2)^2
%+\frac{\lambda}{(\rho-\omega)}(\phi_-^2-|\phi_0|^2)^2+
%\frac{2\lambda}{(\rho^2-\omega^2)} \phi_+^2 \phi_-^2
\end{align}
where $\lambda$ denotes a dimensionless quartic coupling.
Since the valley of the potential~(\ref{powerlaw inflation:V})
corresponds to the inflaton, for a negative $\mu_r$, $\phi_-$ 
corresponds to the inflaton while $\phi_+$ is a BD scalar. 
As we argued above, this choice for $\mu_r$ also provides 
a large parameter space for charge amplification. The BD scalar acquires
a potential, which is however steeper than that of the inflaton. 
This may provide a way to anchoring the BD scalar to a fixed value today
thus helping to make the theory compatible with experiments.

\section{Conclusion}
\label{Conclusion}

 By generalizing Brans-Dicke (BD) theory to a theory with a complex scalar
field~(\ref{our lagrangian}), we have constructed a baryogenesis model
operative during inflation. 
%allows for CP violation in the scalar sector.
In its most general disguise our BD theory contains one physical 
CP violating phase which can be used to amplify the corresponding scalar
current during inflation.

 Our starting assumption is that at the beginning 
of inflation the Universe contains domains of several Hubble
volumes which contain a small amount of initial scalar charge.
During inflation these domains are stretched to a volume
larger than our visible universe, and at the same time the charge density
gets (exponentially) enhanced. 
When inflation ends, CP violation turnes off and the Universe heats up,
increasing the entropy of the Universe by a large factor. 
The scalar charge 
decays in a charge conserving way to ordinary matter through a
tree level coupling, creating a net lepton and/or baryon number.
The lepton minus baryon number is then processed via sphalerons
to the lepton and baryon number of a magnitude which we observe today. 

We have also shown that our mechanism can be incorporated into 
an extended inflationary model with a minimal adaptation
when considering the real and
imaginary part of the field as two independent real fields. The
real part acts like a Brans-Dicke scalar while the imaginary part
is the inflaton. The theory can provide a natural potential
for the Brans-Dicke scalar, potentially anchoring the expectation
value today and loosening the constraints on the theory. More
detailed calculations are needed to further investigate these
proposals and compare them with the experimental bounds. 

The main attractive feature of the generalised Brans-Dicke theory
we consider here is in the fact that a model for baryogenesis 
is implemented in the theory of gravity itself. 
A related model, dubbed gravitational baryogenesis,
was considered in Ref.~\cite{Davoudiasl:2004gf}, where baryons are produced
via a dimension six operator which couples 
a standard model fermionic current to the gradient of the Ricci scalar. 
The authors argue that in homogeneous limit this interaction can be 
thought of as a chemical potential for baryon number of the form, 
$\mu_B \propto \dot R \propto \rho^{3/2}$, which therefore decays 
as $1/a^{6}$ in radiation era and as $1/a^{9/2}$ in matter era.
That means that, unless the Universe starts at very high energy scales 
 in a very asymmetric state, that mechanism does not work.  
Some may consider that as a feature which is not very desirable.
Our mechanism does not suffer from such a problem
since we require only a tiny preinflationary charge density which extends
over several Hubble volumes, and which may as well be generated
by statistical fluctuations in preinflationary primordial chaos.
 
 Another interesting aspect of our model is that baryogenesis takes place
during inflation, a very important era in the evolution of the universe
not much considered in combination with baryogenesis. 
The CP and charge violation in the theory is turned off when the
universe enters radiation era, such that 
after inflation the produced charge freezes in.
It would be of interest to further investigate the model
in a more general Scalar-Tensor theory of
gravity or explore the analog in string theory with the dilaton.

\bibliography{references}
\bibliographystyle{unsrt}

\end{document}